\documentclass[twocolumn]{aastex631}
\usepackage{hyperref}

\newcommand\oiii{[\ion{O}{3}]}
\newcommand\Mgii{\ion{Mg}{2}}
\newcommand\Nii{[\ion{N}{2}]}
\newcommand\Ciii{\ion{C}{3]}}
\newcommand\Ciiii{\ion{C}{4}}
\newcommand\Sii{[\ion{S}{2}]}
\newcommand\Feii{\ion{Fe}{2}}

\newcommand{\ha}{\ensuremath{\mathrm{H\alpha}}}
\newcommand{\hb}{\ensuremath{\mathrm{H\beta}}}

\definecolor{dqcolor}{RGB}{255,0,0}

\usepackage{booktabs}
\usepackage{xcolor}
\usepackage{soul}
\usepackage{textcomp}
\usepackage{hyperref}
\usepackage{upgreek}

\begin{document}

\title{Discovery of Repeating Transitions in 16 Changing-look Active Galactic Nuclei}

\author{Qian Dong}
\affiliation{Department of Astronomy, Xiamen University, Xiamen, Fujian 361005, People's Republic of China}

\author{Zhi-Xiang Zhang}
\affiliation{College of Physics and Information Engineering, Quanzhou Normal University, Quanzhou, Fujian 362000, People's Republic of China}

\author{Wei-Min Gu}
\affiliation{Department of Astronomy, Xiamen University, Xiamen, Fujian 361005, People's Republic of China}

\author{Mouyuan Sun}
\affiliation{Department of Astronomy, Xiamen University, Xiamen, Fujian 361005, People's Republic of China}

\author{Wei-Jian Guo}
\affiliation{Key Laboratory of Optical Astronomy, National Astronomical Observatories, Chinese Academy of Sciences, Beijing 100012, China}

\author{Zhen-Yi Cai}
\affiliation{Department of Astronomy, University of Science and Technology of China, Hefei 230026, People’s Republic of China}

\author{Jun-Xian Wang}
\affiliation{Department of Astronomy, University of Science and Technology of China, Hefei 230026, People’s Republic of China}

\author{Yong-Gang Zheng}
\affiliation{Department of Physics, Yunnan Normal University, Kunming, 650092, People's Republic of China}

\correspondingauthor{Zhi-Xiang Zhang, Wei-Min Gu}
\email{zzx@qztc.edu.cn, guwm@xmu.edu.cn}

\begin{abstract}
The repeating changing-look active galactic nuclei (RCL AGNs) exhibit multiple appearances and disappearances of broad emission lines (BELs), whose underlying mechanism remains a puzzle. Expanding the sample of RCL AGNs is valuable for constraining the transition timescale and probing the accretion physics driving CL behaviors. This study aims to identify RCL AGNs using the multi-epoch spectroscopic data of confirmed CL AGNs from the Sloan Digital Sky Survey, Large Sky Area Multi-Object Fiber Spectroscopic Telescope, and Dark Energy Spectroscopic Instrument, supplemented with mid-infrared (MIR) light curves. Through selection criteria and visual inspection, we identify 22 RCL AGNs among 299 CL AGNs, corresponding to an occurrence rate of about 7\%, indicating that repeated transitions are not extremely rare in CL AGNs. Among the 22 RCL AGNs, 16 are newly identified, which significantly expands the known RCL AGN sample. Based on the spectra and densely sampled MIR light curves, we derive MIR variability timescales for 18 RCL AGNs, and find no significant correlation between the timescale and the black hole mass. 

\end{abstract}

\section{Introduction}\label{sec:Introduction} 
Active galactic nuclei (AGNs) are energetic sources powered by accretion of material onto supermassive black holes (SMBHs) at the centers of galaxies. They are characterized by high luminosity and variability across multiple wavelengths \citep{1997ARA&A..35..445U,2000ApJ...540..652W,2007AJ....134.2236S,2016ApJ...817..119K}. The typical optical spectra of AGNs consist of a power-law continuum together with both broad emission lines (BELs; $1000-20000\,\mathrm{km\,s^{-1}}$) and narrow emission lines (NELs; $300-1000\,\mathrm{km\,s^{-1}}$). AGNs are commonly classified into Type 1 and Type 2 AGNs based on their optical spectral features. Type 1 AGNs exhibit both BELs and NELs, yet Type 2 AGNs only exhibit NELs. Spectropolarimetric observations revealed that some Type 2 AGNs also display hidden BELs similar to those in Type 1 AGNs \citep{1985ApJ...297..621A,1990ApJ...355..456M}. This finding promotes the formulation of the AGN unified model, suggesting that Type 1 and Type 2 AGNs are intrinsically the same objects, and the observed differences primarily arise from the orientation of the dusty torus relative to our line of sight \citep{1995PASP..107..803U}.

Most AGNs exhibit random optical continuum variations of about $0.2\,\rm{mag}$ on timescales of months to years \citep{2005ApJ...633..638W}. The optical continuum is generally produced from the accretion disk, and such modest variability is usually attributed to disk instabilities \citep{1998ApJ...504..671K}. The most widely used accretion disk model is the standard thin disk \citep[SSD;][]{1973A&A....24..337S}, which predicts that the disk should remain stable on the viscous timescale, typically on the order of $10^{4}$ years. AGNs are not expected to show dramatic luminosity changes on much shorter timescales. 

However, a subset of AGNs has shown notable changes in spectra over the past two decades. These objects display the appearance or disappearance of BELs within several months to decades, referred to as turn-on or turn-off transitions, and are classified as changing-look AGNs \citep[CL~AGNs;][]{2015ApJ...800..144L,MacLeod2016}.

To date, approximately $1100$ CL~AGNs have been identified through repeated spectroscopic observations. Most of them are identified using data from the Sloan Digital Sky Survey \citep[SDSS;][]{2000AJ....120.1579Y}, the Dark Energy Spectroscopic Instrument \citep[DESI;][]{2022AJ....164..207D}, and the Large Sky Area Multi-Object Fiber Spectroscopic Telescope \citep[LAMOST;][]{2012RAA....12.1197C}. For example, \citet{2024ApJS..270...26G,2025ApJS..278...28G} confirmed over $700$ CL~AGNs based on SDSS and DESI spectra, \citet{2024ApJ...966...85Z} and \citet{2025arXiv251006753L} confirmed $116$ and $34$ CL~AGNs using repeated SDSS spectroscopy, respectively. \citet{Yang2018} identified $21$ CL~AGNs, $15$ of them identified based on SDSS and LAMOST spectra, while the remaining six objects are confirmed through newly spectroscopic observations from other telescopes. In addition, some CL~AGNs are selected based on significant photometric variability and then confirmed with follow-up spectroscopy, such as \citet{Graham2020} and \citet{2025ApJ...980...91Y}. 

Although the number of CL~AGNs continues to grow, their physical mechanism remains a puzzle \citep{2024SerAJ.209....1K}. The SSD model predicts that significant changes in the accretion rate should happen on timescales of $\sim{10^4\,\rm{yr}}$, which is much longer than the observed CL timescales \citep{2011MNRAS.413.2259S}. This discrepancy poses a challenge to the understanding of traditional accretion theory. Several mechanisms have been proposed to explain the CL phenomenon, including dust obscuration \citep{1989ApJ...340..190G,1992AJ....104.2072T}, tidal disruption event \citep[TDE;][]{1988Natur.333..523R,1989ApJ...346L..13E,2015MNRAS.452...69M}, and changes in the accretion rate \citep{2014MNRAS.438.3340E}. Among these, the transition of the accretion state or the change in the accretion rate are widely supported explanations \citep[e.g.,][]{2015ApJ...800..144L,2016ApJ...826..188R,2016MNRAS.455.1691R,MacLeod2016,Sheng2017,Yang2018,2022ApJ...930...46L,2024ApJS..270...26G,2024ApJ...962..122K,2025ApJ...985..185M}. A number of revised accretion models have been proposed to solve the timescale mismatch, such as the magnetically elevated disc \citep{2019MNRAS.483L..17D}, the magnetic disk outflow \citep{2021ApJ...916...61F,2023ApJ...958..146W,2023MNRAS.526.2331C}, and the radiation instability zone of the disk \citep{2020A&A...641A.167S,2025ApJ...988..207L}. 

With increasing observation data, partial CL~AGNs have been observed to undergo more than one transition, characterized by the disappearance followed by the reappearance of BELs, or vice versa. These sources are referred to as repeating CL~AGNs (RCL~AGNs). Compared to CL~AGNs, RCL~AGNs offer unique advantages for understanding the physical mechanisms of state transitions. Multiple transitions provide tighter constraints on the duration of a full CL cycle and the characteristic timescales of spectral changes. Moreover, their repeating nature provides a valuable laboratory for testing accretion models and exploring the long-term physical processes of CL behavior. Recent studies have attempted to find the potential physical mechanism based on the RCL~AGNs. For example, \cite{2024ApJ...970...85W} found an inverse trend between the black hole mass and CL cycle timescale using multi-epoch spectra of RCL~AGNs. They proposed that CL behavior may be driven by the change in accretion rate due to the gas supply. However, their work was based on only $11$ confirmed RCL~AGNs, and the derived timescales spanned a wide range with large uncertainties. Therefore, expanding the RCL~AGN sample is essential for understanding the physical mechanisms driving CL behavior. 

To date, confirmed RCL~AGNs remain few, with approximately $28$ reported \citep{2024ApJ...970...85W,2025ApJ...981..129W,2025A&A...693A.173L,2025ApJ...986..160D}. Discovering more RCL~AGNs will be crucial for clarifying and understanding the physical processes that drive their transitions. In this study, we aim to identify RCL~AGNs among previously confirmed CL~AGNs. We first compile CL~AGNs reported in the literature and collect their multi-epoch spectroscopic data from the SDSS, LAMOST, and DESI surveys. The identification of RCL behavior focuses on examining whether a previously known CL~AGN exhibits additional CL transitions in multi-epoch spectra, which we assess through a combination of selection criteria and careful visual inspection. Through this process, we identify a sample of $22$ RCL~AGNs, of which $16$ are newly confirmed.

The content of this paper is organized as follows. The collection of the CL~AGN sample, spectroscopic data, and the selection criteria for identifying RCL~AGNs are described in Section \ref{sec:data}. We present the results and discussion in Section \ref{sec:discussion}, and summary in Section \ref{sec:summary}. Throughout this paper, we adopt a flat $\mathrm{\Lambda CDM}$ cosmology with $H_0 = 67.66\,\mathrm{km\,s^{-1}\,{Mpc}^{-1}}$ and $\mathrm{\Omega_m} = 0.30966$ \citep{2020A&A...641A...6P}.
 
\section{Sample and data}\label{sec:data}
\subsection{CL~AGN sample}\label{sec: CLAGN sample}

The initial sample used to identify RCL~AGNs in this study is compiled from previously reported CL~AGNs. These CL~AGNs were collected from catalogs published in the following works:
\begin{itemize}
\item $592$ CL~AGNs from \cite{2025ApJS..278...28G};
\item $130$ CL~AGNs from \cite{2024ApJS..270...26G};
\item $116$ CL~AGNs from \cite{2024ApJ...966...85Z};
\item $110$ CL~AGNs from \cite{Graham2020};
\item $82$ CL~AGNs from \cite{2025ApJ...980...91Y};
\item $51$ CL~AGNs from \cite{2025ApJ...986..160D};
\item $34$ CL~AGNs from \cite{2025arXiv251006753L}
\item $27$ CL~AGNs from \cite{2022MNRAS.511...54H};
\item $21$ CL~AGNs each from \cite{Yang2018} and \cite{2024ApJ...966..128W};
\item $19$ CL~AGNs from \cite{Green2022};
\item $17$ CL~AGNs from \cite{2019ApJ...874....8M};
\item $10$ CL~AGNs from \cite{MacLeod2016};
\item Others from the following works: $8$ CL~AGNs from \cite{2023MNRAS.518.2938T}; $7$ CL~AGNs from \cite{2016ApJ...821...33R}; $6$ CL~AGNs from \cite{2019ApJ...883...31F}; $4$ CL~AGNs from \cite{2020MNRAS.497..192H}; $3$ CL~AGNs each from \cite{2019ApJ...883L..44G} and \cite{2020MNRAS.498.2339R}; $2$ CL~AGNs each from \cite{2016ApJ...826..188R} and \cite{2019ApJ...887...15W}; $1$ CL~AGN each from \cite{2017ApJ...835..144G}, \cite{2015ApJ...800..144L}, \cite{2016MNRAS.455.1691R}, and \cite{2018ApJ...858...49W}.
\end{itemize}
We collected $1,269$ CL~AGNs from these works and obtained a final sample of $1,154$ CL~AGNs after removing duplicates. The redshifts of the final CL~AGN sample range from $0.0054$ to $3.3657$.

\subsection{Spectroscopic data}
We collected spectroscopic data for these sources from SDSS, LAMOST, and DESI, along with spectral figures presented in the relevant literature. To identify RCL~AGNs, we required each target to have at least three spectroscopic observations. In the following, we briefly describe the SDSS, LAMOST, and DESI surveys. Information on other telescopes used in the reported studies will not be discussed in detail here.

The SDSS has been conducting spectroscopic observations since 1998, providing a continuous temporal baseline \citep{2000AJ....120.1579Y}. The original SDSS spectrographs were upgraded for the Baryon Oscillation Spectroscopic Survey \citep[BOSS;][]{2013AJ....145...10D}, which extended the wavelength coverage from $3800-9100$\,\AA\ to $3600-10400$\,\AA\ and maintained a spectral resolution of $R\approx2000$ \citep{2008ApJS..175..297A, 2013AJ....145...10D}. In this work, we use high-quality spectra ($\mathrm{zWarning}=0$ or $16$) classified as quasars (`QSO') or galaxies (`GALAXY') from SDSS Data Release 18 \citep[DR18; see][]{2023ApJS..267...44A}, yielding a total of $3,848,269$ available spectra.

The LAMOST began its spectroscopic survey in 2011 and is capable of obtaining up to $4000$ spectra simultaneously \citep{2012RAA....12.1197C,2012RAA....12..723Z}. The low-resolution survey of LAMOST provides wavelength coverage from $3700$ to $9000$\,\AA\ with a resolution of $R\sim1800$. For our analysis, we use spectra from Data Release 12 (DR12) with reliable redshift measurements ($z\neq-9999$). Among them, $320,287$ spectra are classified as `GALAXY' or `QSO'.

The DESI began survey validation (SV) in December 2020 and launched its five-year spectroscopic survey in May 2021 \citep{2024AJ....168...58D}. The first public data release (DR1), issued in March 2025, includes both the SV phase (December 2020–April 2021) and the early main survey (May 2021–June 2022; see \citealt{2025arXiv250314745D}). For this work, we use the DESI summary catalog of AGN and QSO, which contains $16,360,818$ spectra with $\mathrm{ZWARN}=0$.

We cross-matched the CL~AGN sample described in Section \ref{sec: CLAGN sample} with the SDSS, LAMOST, and DESI catalogs using a matching radius of $2{\arcsec}$. In total, $491$ CL~AGNs have at least three spectroscopic observations, including public data (SDSS, LAMOST, and DESI) and follow-up observations in the literature. 

\subsection{Mid-infrared data}\label{sec:MIR_data} 
Photometric light curves provide useful information to examine signatures of CL behavior in AGNs. Many previous studies have used photometric variability to identify or jointly confirm CL candidates \citep[e.g.,][]{MacLeod2016,Yang2018,Graham2020}. Although optical bands have been widely employed, their measurements can be significantly affected by host-galaxy contamination and dust extinction, which dilutes the intrinsic AGN signal \citep[e.g., see][]{Shen2011,Stern2012}. Compared with optical bands, mid-infrared (MIR) observations are far less affected by dust extinction, providing a cleaner trace of intrinsic accretion changes. The MIR emission originates from the dust torus heated by radiation from the accretion disk and thus responds to variations in the central engine with a time delay \citep{2019ApJ...886...33L,2020ApJ...900...58Y,2023MNRAS.522.3439C}. Indeed, MIR variability has already been successfully applied to the identification of CL quasars \citep[e.g.,][]{Jun2015,Sheng2017,Stern2018}. Motivated by these advantages, we adopt MIR photometric data to probe the long-term variability of our sample.

The Wide-field Infrared Survey Explorer (WISE; \citealt{2010AJ....140.1868W}) is a space telescope equipped with a $40\,\rm{cm}$ primary mirror, observing in four MIR bands: W1 ($3.4\,\rm{\upmu m}$), W2 ($4.6\,\rm{\upmu m}$), W3 ($12\,\rm{\upmu m}$), and W4 ($22\,\rm{\upmu m}$). The corresponding sensitivity limits are $16.5$, $15.5$, $11.2$, and $7.9\,\rm{mag}$, respectively. WISE has a 47\arcmin\ field of view and shifts the scanning direction approximately $4^\prime$ per orbit, providing roughly $12$ repeated exposures of a given source per sky pass. The mission began on 14 January 2010 and completed its first full-sky survey within six months. After the exhaustion of its hydrogen coolant, only the W1 and W2 bands remained operational, and the telescope entered hibernation following the third full-sky scan. In September 2013, the telescope was reactivated as the Near-Earth Object WISE (NEOWISE) mission, resuming routine observations in the W1 and W2 bands \citep{2011ApJ...731...53M,2014ApJ...792...30M}.

For this study, we collect W1 and W2 photometry from both the WISE and NEOWISE missions. We exclude data points with photometric uncertainties greater than $0.3\,\rm{mag}$. For each epoch, we compute the mean value of the $\sim12$ repeated exposures. The resulting light curves are then smoothed using the LOWESS method \citep{Cleveland1979}, and outliers deviating by more than $3\sigma$ from the smoothed trend are removed.

\subsection{Spectra fitting and flux calibration}\label{sec_fitting}
\begin{figure}
  \begin{center}
    \resizebox{88mm}{!} {\includegraphics *{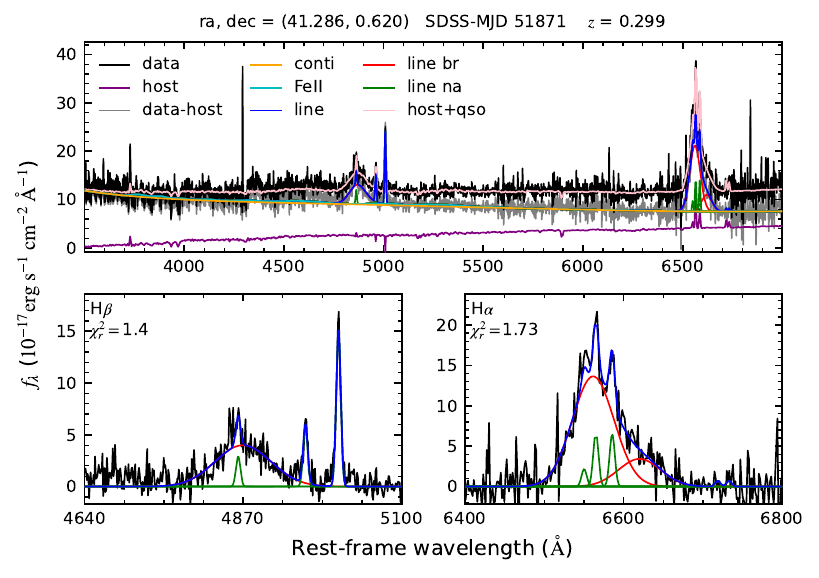}}
    \resizebox{88mm}{!} {\includegraphics *{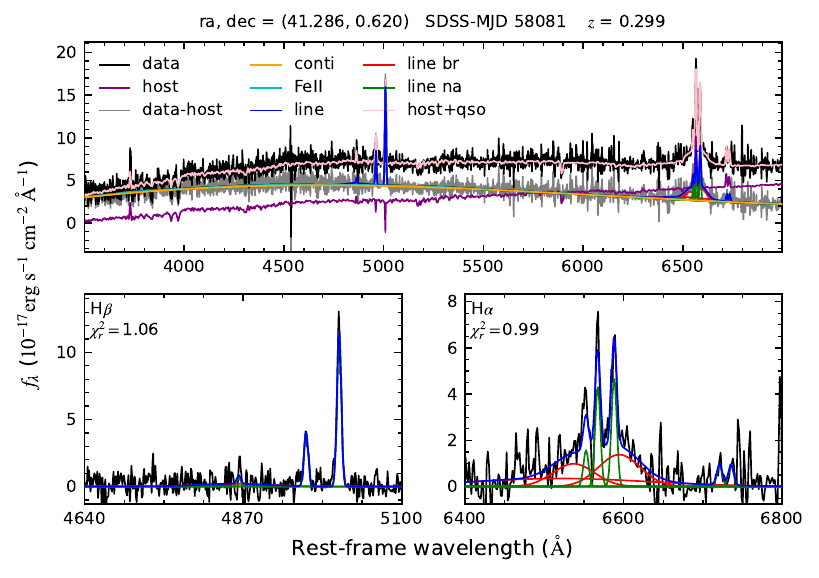}}
    \resizebox{88mm}{!} {\includegraphics *{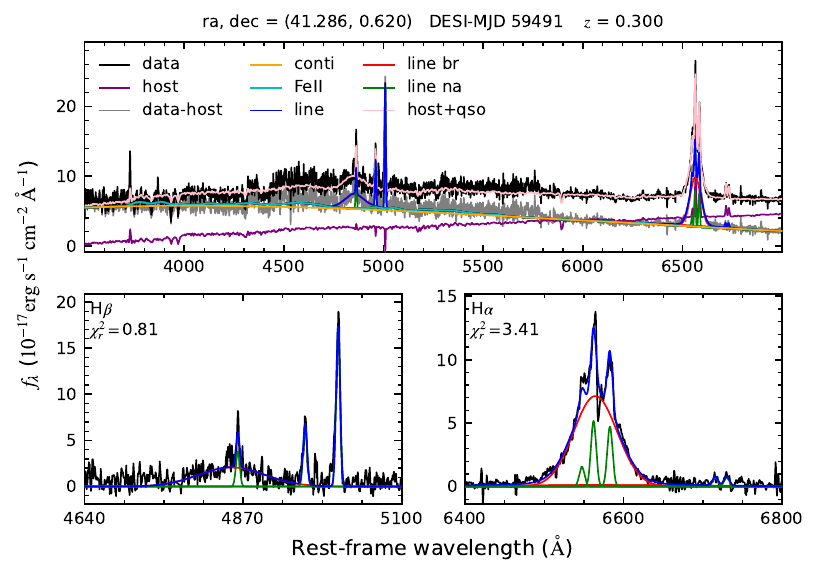}}
\caption{Spectral fitting results for the RCL AGN J0245+0037 at three epochs (MJD 51871, 58081, and 59491). The host galaxy component is determined from the dim-state spectrum and fixed in the fitting of all epochs. The top panel shows the observed spectrum (black), which is divided into the host galaxy (purple), \Feii\ emission, power-law continuum (orange), and emission lines (blue). The emission lines are further decomposed into NELs (green) and BELs (red), which expressed as `na’ and `br’. The second left and right panels present the fitting results of \hb\ and \ha\ emission lines, respectively. From top to bottom, the panels correspond to the three epochs in chronological order. }
\label{fig:J0245+0037}
\end{center}
\end{figure}

We fit the spectra using the public \texttt{QSOFITMORE} package \citep{2021zndo...5810042F}, which is modified on the basis of \texttt{PyQSOFit} \citep{2018ascl.soft09008G}. This package first performs an extinction correction to account for dust reddening effects of galactic. The extinction corrected spectrum is then decomposed into host galaxy and quasar components through principal component analysis (PCA). CL~AGNs provide a unique advantage for spectral decomposition because multi-epoch spectra are available in both bright and dim states. In the dim state, the AGN continuum is significantly weakened, enabling a more reliable constraint on the host galaxy component. Therefore, we first fit the dim-state spectrum to determine the host contribution \citep{2025ApJ...980...91Y}, and then fix this component when fitting the bright-state spectra, allowing only the quasar components to vary, as shown in Figure \ref{fig:J0245+0037}. For the quasar component, the quasar continuum is modeled using a combination of a power-law and \Feii\ template, and the remaining emission lines are fitted with Gaussian functions. Emission lines are decomposed into broad and narrow components based on the FWHM threshold of $1200\,\rm {km\,s^{-1}}$. In the \ha\ complex ($6400-6800 \,\rm{\AA}$), the broad \ha\ component is modeled with three Gaussian profiles, while the narrow \ha, \Nii$\lambda6549$, \Nii$\lambda6585$, \Sii$\lambda6718$, and \Sii$\lambda6732$ are each fitted with a single Gaussian profile. In the \hb\ complex ($4640-5100 \,\rm{\AA}$), the broad \hb\ component is modeled with three Gaussian profiles, the narrow \hb\ with a single Gaussian, and each of the \oiii$\lambda\lambda4959,5007$ doublet is decomposed into a core and a wing component, each modeled with a single Gaussian. For \Mgii\ ($2700–2900 \,\rm{\AA}$), the broad component is modeled with two Gaussians and the narrow component with a single Gaussian. The \Ciii\ ($1850-1970\,\rm{\AA}$) and \Ciiii\ ($1500-1600\,\rm{\AA}$) lines are modeled with two Gaussian profiles. Through this process, \texttt{QSOFITMORE} provides reliable measurements of the BEL properties \citep{2022ApJS..261...32F}.

To ensure the reliability and comparability of flux measurements among the SDSS, LAMOST, and DESI spectra, we adopt different strategies for the LAMOST and DESI data.

For the DESI spectra, an absolute flux calibration is already provided, and we therefore focus on validating its reliability. Following \citet{2025ApJS..278...28G}, we perform two independent checks based on photometric data and the \oiii$\lambda5007$ emission line. First, we use nearly contemporaneous ZTF photometry \citep{2019PASP..131a8002B}. Outliers are removed with a $3\sigma$ clipping procedure implemented in \textsc{Lightkurve} \citep{2018ascl.soft12013L}, and the reference magnitude is taken as the mean ZTF magnitude within $\pm5$ days of the spectroscopic epoch. Synthetic magnitudes are computed from the DESI spectra using \textsc{PyPhot}\footnote{\url{https://mfouesneau.github.io/pyphot/}}, and spectra with differences larger than $0.5\,\mathrm{mag}$ are excluded. Second, we compare the \oiii$\lambda5007$ flux measured from DESI and SDSS spectra. Differences within $20\%$ are considered acceptable, allowing for aperture effects due to the smaller DESI fiber ($1.5\arcsec$; see \citealt{2024AJ....168...58D}) relative to SDSS. Spectra with larger discrepancies are discarded. When both ZTF and \oiii\ information are available, both criteria must be satisfied; otherwise, only the available criterion is applied. Spectra lacking both constraints are excluded.

For the LAMOST spectra, the pipeline provides a reliable spectral shape but does not deliver an absolute flux calibration. Therefore, an additional scaling is required before comparing emission-line variability. We rescale the LAMOST spectra using the \oiii$\lambda5007$ emission line, adopting the SDSS spectra as the reference. The \oiii\ line originates from the extended narrow-line region and is generally expected to remain stable on decadal timescales \citep{1992PASP..104..700V,2017PASP..129b4007F}. Moreover, the fiber sizes of SDSS ($3\arcsec$; see \citealt{2008ApJS..175..297A}) and LAMOST ($3.3\arcsec$; see \citealt{2012RAA....12.1197C}) are comparable, so aperture effects are expected to be minor. However, differences in the seeing between SDSS and LAMOST may introduce additional uncertainties into the spectral calibration. To evaluate the reliability of the \oiii-based calibration using the SDSS spectra as the reference, we derive calibration factors from nearly contemporaneous ZTF photometry and compare them with those obtained from the \oiii\ flux scaling. This comparison is performed for $30$ CL~AGNs that have SDSS spectra, LAMOST spectra, and suitable nearly contemporaneous ZTF observations. We find that approximately 90\% of these sources show consistency within 20\%, suggesting that the systematic uncertainty introduced by aperture and seeing differences is generally moderate for most objects in our sample. Therefore, the \oiii-based calibration using SDSS spectra is expected to be reliable for the majority of the LAMOST spectra in this work. The scaling factor is defined as the ratio of the \oiii\ flux measured in the SDSS spectrum to that in the LAMOST spectrum, and is applied to rescale the LAMOST spectrum. This approach is applicable to sources at $z \leq 0.83$, for which the \oiii$\lambda5007$ line is covered by LAMOST; higher-redshift spectra without \oiii\ coverage are excluded.  

\subsection{Selection of RCL~AGNs}
\subsubsection{Criterion of selection}
We identify RCL~AGNs based on significant variations in their BELs. 
For sources with $z < 0.85$, we focus on the \ha, \hb, and \ion{Mg}{2} lines, while for $z \geq 0.85$, we examine \ion{Mg}{2}, \Ciii, and \Ciiii\ lines. To exclude cases with only marginal BEL variability, we adopt the selection criterion proposed by \cite{2025ApJ...986..160D}, which requires a substantial relative flux change between the bright and dim states. The relative variation, $R_{\rm s}$, is defined as:
\begin{equation}
R_{\mathrm{s}} = \frac{S_{\mathrm{b}} - S_{\mathrm{d}}}{S_{\mathrm{b}}},
\end{equation}
where $S_{\mathrm{b}}$ and $S_{\mathrm{d}}$ are the integrated BEL fluxes in the bright and dim states, respectively. A source is considered to exhibit a significant BEL variation when $R_{\rm s} > 0.3$.

The spectra used for the previous CL identification are denoted as $\rm{T1}$ (earlier epoch) and $\rm{T2}$ (later epoch). In our collected sample, $331$ objects have at least one additional spectroscopic observation either before $\rm{T1}$ or after $\rm{T2}$. After excluding LAMOST spectra lacking \oiii$\lambda5007$ coverage and DESI spectra failing the validation criteria described in Section~\ref{sec_fitting}, $299$ sources retain at least one reliable additional spectrum, providing an opportunity to search for RCL behavior. Among them, $248$ objects have all three spectra publicly available, allowing quantitative spectral fitting and selection. Of these, $121$ objects satisfy our selection criteria. For the remaining $51$ objects, some of the spectra are unavailable, preventing quantitative analysis. In these cases, we rely on published spectral figures in the literature to visually inspect the BELs across different epochs and to assess their variability.

\subsubsection{Visual inspection}
\label{sec:visual}
Although the selection criterion based on fitted spectral parameters provides an efficient and objective way to identify candidate RCL~AGNs, such an approach is not sufficient on its own. Automated fitting can be affected by spectral quality, continuum modeling, or line-blending issues, which may lead to false positives or missed candidates. To ensure the reliability of our sample, we therefore perform an additional step of visual inspection. This second round of selection allows us to verify the BEL variability suggested by the fitting results, to exclude spurious detections caused by poor data quality, and to recover genuine transitions that may not be fully captured by the fitting procedure. In this way, the combination of quantitative fitting and visual inspection ensures both efficiency and robustness in our identification of RCL~AGNs.

The visual inspection focuses on three aspects: (1) for low redshift objects, the spectra in the dim state should exhibit clear \oiii\ emission lines but no detectable BELs, thereby ensuring both spectral quality and a genuine transition toward a Type~2 AGN; (2) the appearance or disappearance of the BELs should be consistent with brightening or dimming trends observed in the MIR light curves; (3) spectra with unreliable fitting results are excluded. 

Applying this procedure, we identify $20$ RCL~AGNs from the $121$ sources with quantified BEL variations. Among them, J1638+2827 shows no significant MIR variability, likely because its MIR brightness approaches the WISE sensitivity limit. Nevertheless, it exhibits prominent BEL changes in multi-epoch spectra, and we therefore retain them as RCL~AGNs based on their clear spectroscopic transitions. The remaining sources (such as J0023+0035, see Figure~\ref{Fig_RCLAGN}) are brighter in the MIR band and are not limited by the observation sensitivity. They show clear changes in multi-epoch spectra and MIR light curves. In addition, J1053+3024 exhibits the undetectable broad \hb\ line in the spectrum at MJD 58141, and it clearly reappears at MJD 58514. Although the recovered broad line remains relatively weak, the MIR emission continues to brighten after MJD~58514, suggesting that the transition may still be ongoing. We thus retain this source in the final RCL~AGN sample.

For the other $51$ sources only spectral figures are available in the literature, we visually inspect the published spectra figures and find two additional objects that show significant repeating BEL changes. Their RCL behaviors are described in Section \ref{sec:absence}. In total, we identify $22$ RCL~AGNs through the combined use of quantitative selection and visual inspection. These sources are listed in Table~\ref{tab:RCLAGNs}, and their multi-epoch spectral transitions are presented in Figure~\ref{Fig_RCLAGN}. 

\subsubsection{RCL~AGNs identified by visual inspection only}\label{sec:absence}
As what we said above, in addition to the sources selected through quantitative BEL measurements, our final RCL~AGN sample includes two objects that could not be assessed using flux-based criteria. For these targets, only two spectra are available in public archives, and the third epoch comes from spectra presented in the literature. Consequently, quantitative evaluation of BEL variations is not possible, and their identification relies solely on visual inspection of multi-epoch spectra. The details of these two sources are as follows:

\begin{itemize}
\item \textbf{J0745+3809}: This source was identified as a turn-off CL~AGN by \citet{2019ApJ...874....8M} based on SDSS (MJD=51873) and MMT (MJD=58081) observations (see Figure 2 of \citealt{2019ApJ...874....8M}). The recent DESI spectrum (MJD=59509) shows the reappearance of the \hb\ BEL, indicating a subsequent turn-on transition and confirming this object as an RCL~AGN.
\item \textbf{J0927+0433}: This object was previously reported as a turn-off CL~AGN by \citet{2019ApJ...874....8M}. The SDSS spectrum (MJD=52254) exhibits a clear \hb\ BEL, while the MMT spectrum (MJD=57819) shows its absence (see Figure 2 of \citealt{2019ApJ...874....8M}). Subsequent observations from LAMOST (MJD=59193) show a weak \hb\ BEL, and the latest DESI spectrum (MJD=59622) reveals a strong reappearance of \hb, confirming its RCL behavior.
\end{itemize}

For these objects, although they cannot be selected based on measured BEL flux variations, their multi-epoch spectra provide clear evidence of RCL transitions, justifying their inclusion in our final sample. To visually illustrate these transitions, we reconstruct these spectra from the published figures in \citet{2019ApJ...874....8M} using \texttt{WebPlotDigitizer}\footnote{\url{https://automeris.io/wpd/}}. The reconstructed spectra and the corresponding light curves are presented in Figure~\ref{Fig_RCLAGN}. Owing to the limited resolution of the reconstructed spectra, they are used solely for visualization and are excluded from all quantitative analyses, including spectral fitting and flux calibration.

\begin{deluxetable*}{ccccccccccccccccc}
% \tabletypesize{\scriptsize}
\tablewidth{0.24pt}
\setlength{\tabcolsep}{1.5pt}
\tabletypesize{\scriptsize} 
\tablecaption{RCL~AGN sample}
\centering
\tablehead{\colhead{Name} & \colhead{R.A.} & \colhead{Dec.}& \colhead{z}& \colhead{ $\mathrm{MJD_1}$ }& \colhead{ $\mathrm{MJD_2}$ } & \colhead{$\mathrm{MJD_3}$} & \colhead{$\log M_{\mathrm{BH}}$($M_{\odot}$)}& \colhead{$\log \lambda_{\mathrm{Edd,1}}$}& \colhead{ $\log \lambda_{\mathrm{Edd,2}}$} & \colhead{ $\log \lambda_{\mathrm{Edd,3}}$ } & \colhead{Timescale($\rm{yr}$)} & \colhead{ $R_{s,1}$} & \colhead{$R_{s,2}$}
& \colhead{Line} & \colhead{Ref}  \\
 \colhead{(1)} &  \colhead{(2)} &  \colhead{(3)} &  \colhead{(4)} &  \colhead{(5)} &  \colhead{(6)} &  \colhead{(7)} &   \colhead{(8)} &  \colhead{(9)} &  \colhead{(10)}&  \colhead{(11)}&  \colhead{(12)} &  \colhead{(13)}&  \colhead{(14)} &  \colhead{(15)} &  \colhead{(16)}}
\startdata
J0043+1344	&	10.91383	&	13.74347	&	0.527	&	51879	&	58462	&	59549	&	8.98 $\pm$ 0.04	&	-1.81 $\pm$ 0.04	&	-2.31 $\pm$ 0.04	&	-1.90 $\pm$ 0.04	&	2.33	&	0.85	&	0.76	&	\Mgii	&	(a) \\
&	 	&	 	&	 	&	 	&	 	&	  &	 	&	 	&	 &	 	&	 	&	 	0.89	&	0.85  &	 \hb	&	  \\
J0205-0456	&	31.31158	&	-4.94437	&	0.363	&	55944	&	57723	&	59530	&	8.32 $\pm$ 0.06	&	-1.52 $\pm$ 0.06	&	-1.89 $\pm$ 0.06	&	-1.56 $\pm$ 0.06	&	4.07	&	0.75	&	0.63	&	\hb	&	(b) \\
J0207-0609	&	31.93359	&	-6.16556	&	0.650	&	55924	&	56660	&	59525	&	9.08 $\pm$ 0.04	&	-1.84 $\pm$ 0.04	&	-1.75 $\pm$ 0.04	&	-2.19 $\pm$ 0.04	&	3.60	&	0.50	&	0.66	&	\hb	&	(c) \\
J0245+0037	&	41.28614	&	0.61966	&	0.299	&	51871	&	58081	&	59491	&	8.50 $\pm$ 0.07	&	-1.53 $\pm$ 0.07	&	-1.85 $\pm$ 0.07	&	-1.79 $\pm$ 0.07	&	4.19	&	0.89	&	0.83	&	\hb	&	(b) \\
J0726+4101	&	111.73366	&	41.02668	&	0.130	&	55205	&	56193	&	59619	&	8.58 $\pm$ 0.04	&	-2.07 $\pm$ 0.04	&	-2.36 $\pm$ 0.04	&	-2.06 $\pm$ 0.04	&	2.15	&	0.69	&	0.72	&	\hb	&	(c) \\
J0745+3809	&	116.29992	&	38.15314	&	0.237	&	51873	&	58081	&	59509	&	8.85 $\pm$ 0.03	&	-1.90 $\pm$ 0.03	&	...	&	-2.09 $\pm$ 0.03	&	4.04	&	...	&	...	&	\hb	&	(a) \\
J0927+0433	&	141.75958	&	4.55228	&	0.322	&	52254	&	57819	&	59622	&	8.92 $\pm$ 0.04	&	-1.82 $\pm$ 0.04	&	...	&	-2.02 $\pm$ 0.04	&	2.70	&	...	&	...	&	\hb	&	(a) \\
J0938+0743	&	144.55113	&	7.72778	&	0.022	&	52733	&	54804	&	59551	&	7.86 $\pm$ 0.08	&	-2.39 $\pm$ 0.08	&	-3.46 $\pm$ 0.08	&	-2.33 $\pm$ 0.08	&	3.36	&	0.84	&	0.93	&	\hb	&	(d) \\
&	 	&	 	&	 	&	 	&	 	&	  &	 	&	 	&	 &	 	&	 	&	 	0.82	&	0.91	&	 \ha	&	  \\
J1028+2351	&	157.09521	&	23.85716	&	0.174	&	53734	&	57103	&	58937	&	8.29 $\pm$ 0.05	&	-1.80 $\pm$ 0.05	&	-2.09 $\pm$ 0.09	&	-1.89 $\pm$ 0.05	&	4.23	&	0.97	&	0.75	&	\hb	&	(e) \\
J1053+3024	&	163.35587	&	30.40538	&	0.249	&	53463	&	58141	&	58514	&	8.76 $\pm$ 0.03	&	-1.96 $\pm$ 0.03	&	-2.23 $\pm$ 0.03	&	-2.21 $\pm$ 0.03	&	2.86	&	0.70	&	0.43	&	\hb	&	(b) \\
J1118+3203	&	169.62350	&	32.06664	&	0.365	&	53431	&	56367	&	59697	&	8.72 $\pm$ 0.03	&	-1.98 $\pm$ 0.03	&	-2.63 $\pm$ 0.04	&	-2.29 $\pm$ 0.04	&	2.19	&	0.63	&	0.66	&	\hb	&	(f) \\
J1412+5400	&	213.22466	&	54.00399	&	0.187	&	52762	&	58561	&	59738	&	7.80 $\pm$ 0.10	&	-1.74 $\pm$ 0.10	&	-2.78 $\pm$ 0.13	&	-1.68 $\pm$ 0.10	&	3.37	&	0.44	&	0.66	&	\hb	&	(c) \\
J1533+0110	&	233.48329	&	1.17492	&	0.143	&	51641	&	54561	&	59643	&	8.01 $\pm$ 0.14	&	-2.22 $\pm$ 0.14	&	-2.13 $\pm$ 0.14	&	-2.39 $\pm$ 0.14	&	6.96	&	0.74	&	0.78	&	\hb 	&	(f) \\
&	 	&	 	&	 	&	 	&	 	&	  &	 	&	 	&	 &	 	&	 	&	 	0.90	&	0.94	&	 \ha	&	  \\
J1638+2827	&	249.72060	&	28.45220	&	2.188	&	54553	&	55832	&	59377	&	...	&	...	&	...	&	...	&	...	&	0.94	&	0.92	&	\Ciiii	&	(g) \\
J2146+0032	&	326.71231	&	0.54277	&	0.335	&	53227	&	55478	&	59486	&	8.56 $\pm$ 0.04	&	-2.09 $\pm$ 0.04	&	-2.31 $\pm$ 0.04	&	-2.16 $\pm$ 0.05	&	3.34	&	0.79	&	0.80	&	\ha	&	(c) \\
J2252+0109	&	343.16821	&	1.16631	&	0.534	&	52178	&	55500	&	59494	&	9.02 $\pm$ 0.06	&	-2.20 $\pm$ 0.06	&	-1.79 $\pm$ 0.06	&	-2.18 $\pm$ 0.06	&	3.26	&	0.55	&	0.75	&	\hb	&	(h) \\ \hline
J0023+0035	&	5.79605	&	0.58820	&	0.422	&	55480	&	58069	&	59493	&	9.17 $\pm$ 0.04	&	-1.70 $\pm$ 0.04	&	-2.18 $\pm$ 0.04	&	-1.91 $\pm$ 0.04	&	3.12	&	0.79	&	0.68	&	\hb	&	(b) \\
J1011+5442	&	152.97075	&	54.70178	&	0.246	&	52652	&	57073	&	59601	&	7.94 $\pm$ 0.07	&	-1.03 $\pm$ 0.07	&	-1.51 $\pm$ 0.07	&	-1.24 $\pm$ 0.07	&	4.47	&	1.00	&	1.00	&	\hb	&	(i) \\
&	 	&	 	&	 	&	 	&	 	&	  &	 	&	 	&	 &	 	&	 	&	 	0.95	&	0.93	&	 \ha	&	  \\
J1104+6343	&	166.09671	&	63.71814	&	0.164	&	52370	&	54498	&	59600	&	8.29 $\pm$ 0.05	&	-2.26 $\pm$ 0.05	&	-2.47 $\pm$ 0.05	&	-2.26 $\pm$ 0.05	&	4.27	&	1.00	&	1.00	&	\hb	&	(f) \\
J1324+2802	&	201.09158	&	28.04398	&	0.124	&	53471	&	56718	&	57755	&	8.15 $\pm$ 0.06	&	-2.35 $\pm$ 0.06	&	-2.67 $\pm$ 0.15	&	-1.83 $\pm$ 0.06	&	2.66	&	0.90	&	0.90	&	\hb	&	(e) \\
J1324+4802	&	201.23871	&	48.04478	&	0.272	&	52759	&	56805	&	57871	&	8.60 $\pm$ 0.04	&	-1.74 $\pm$ 0.04	&	-2.42 $\pm$ 0.04	&	-1.67 $\pm$ 0.04	&	2.69	&	0.73	&	0.52	&	\hb	&	(h) \\
J1617+0638	&	244.29758	&	6.64264	&	0.229	&	53501	&	56776	&	59692	&	8.94 $\pm$ 0.03	&	-1.84 $\pm$ 0.04	&	-2.09 $\pm$ 0.03	&	-2.08 $\pm$ 0.03	&	5.30	&	0.99	&	0.99	&	\hb	&	(e) \\
\enddata
\tablenotetext{}{NOTE—Columns: (1) Source name, (2) right ascension, (3) declination, (4) redshift, (5) MJD of the representative spectrum for the first state, (6) MJD of the representative spectrum for the second state, (7) MJD of the representative spectrum for the third state, (8) black hole mass, (9) the Eddington ratio estimated from the first epoch, (10) the Eddington ratio estimated from the second epoch, (11) the Eddington ratio estimated from the third epoch, (12) MIR variability timescale during the CL period in the rest frame, (13) relative variation of BELs between the spectral taken at $\mathrm{MJD}_1$ and $\mathrm{MJD}_2$, (14) relative variation of BELs between the spectral taken at $\mathrm{MJD}_2$ and $\mathrm{MJD}_3$, (15) the emission lines that show significant variations across multiple observations, (16) the CL~AGNs has reported by previous study: (a)\cite{2019ApJ...874....8M}, (b)\cite{Green2022}, (c)\cite{2025ApJS..278...28G}, (d)\cite{2016ApJ...821...33R}, (e)\cite{2025ApJ...986..160D}, (f)\cite{Yang2018}, (g)\cite{2024ApJS..270...26G}, (h)\cite{MacLeod2016}, (i)\cite{2016MNRAS.455.1691R}.
}
\label{tab:RCLAGNs}
\end{deluxetable*}

\noprint{\figsetstart}
\noprint{\figsetnum{2}}
\noprint{\figsettitle{The spectra and light curves of RCL AGNs.}}

\figsetgrpstart
\figsetgrpnum{2.1}
\figsetgrptitle{RCL AGN_J0043+1344}
\figsetplot{Fig_J0043+1344.pdf}
\figsetgrpnote{The left panel shows the spectra of the RCL AGN J0043+1344, and the right panel shows the light curves in the W1 and W2 bands.}
\figsetgrpend

\figsetgrpstart
\figsetgrpnum{2.2}
\figsetgrptitle{RCL AGN_J0205-0456}
\figsetplot{Fig_J0205-0456.pdf}
\figsetgrpnote{The left panel shows the spectra of the RCL AGN J0205-0456, and the right panel shows the light curves in the W1 and W2 bands.}
\figsetgrpend

\figsetgrpstart
\figsetgrpnum{2.3}
\figsetgrptitle{RCL AGN_J0207-0609}
\figsetplot{Fig_J0207-0609.pdf}
\figsetgrpnote{The left panel shows the spectra of the RCL AGN J0207-0609, and the right panel shows the light curves in the W1 and W2 bands.}
\figsetgrpend

\figsetgrpstart
\figsetgrpnum{2.4}
\figsetgrptitle{RCL AGN_J0245+0037}
\figsetplot{Fig_J0245+0037.pdf}
\figsetgrpnote{The left panel shows the spectra of the RCL AGN J0245+0037, and the right panel shows the light curves in the W1 and W2 bands.}
\figsetgrpend

\figsetgrpstart
\figsetgrpnum{2.5}
\figsetgrptitle{RCL AGN_J0726+4101}
\figsetplot{Fig_J0726+4101.pdf}
\figsetgrpnote{The left panel shows the spectra of the RCL AGN J0726+4101, and the right panel shows the light curves in the W1 and W2 bands.}
\figsetgrpend

\figsetgrpstart
\figsetgrpnum{2.6}
\figsetgrptitle{RCL AGN_J0745+3809}
\figsetplot{Fig_J0745+3809.pdf}
\figsetgrpnote{The left panel shows the spectra of the RCL AGN J0745+3809, and the right panel shows the light curves in the W1 and W2 bands.}
\figsetgrpend

\figsetgrpstart
\figsetgrpnum{2.7}
\figsetgrptitle{RCL AGN_J0927+0433}
\figsetplot{Fig_J0927+0433.pdf}
\figsetgrpnote{The left panel shows the spectra of the RCL AGN J0927+0433, and the right panel shows the light curves in the W1 and W2 bands.}
\figsetgrpend

\figsetgrpstart
\figsetgrpnum{2.8}
\figsetgrptitle{RCL AGN_J0938+0743}
\figsetplot{Fig_J0938+0743.pdf}
\figsetgrpnote{The left panel shows the spectra of the RCL AGN J0938+0743, and the right panel shows the light curves in the W1 and W2 bands.}
\figsetgrpend

\figsetgrpstart
\figsetgrpnum{2.9}
\figsetgrptitle{RCL AGN_J1028+2351}
\figsetplot{Fig_J1028+2351.pdf}
\figsetgrpnote{The left panel shows the spectra of the RCL AGN J1028+2351, and the right panel shows the light curves in the W1 and W2 bands.}
\figsetgrpend

\figsetgrpstart
\figsetgrpnum{2.10}
\figsetgrptitle{RCL AGN_J1053+3024}
\figsetplot{Fig_J1053+3024.pdf}
\figsetgrpnote{The left panel shows the spectra of the RCL AGN J1053+3024, and the right panel shows the light curves in the W1 and W2 bands.}
\figsetgrpend

\figsetgrpstart
\figsetgrpnum{2.11}
\figsetgrptitle{RCL AGN_J1118+3203}
\figsetplot{Fig_J1118+3203.pdf}
\figsetgrpnote{The left panel shows the spectra of the RCL AGN J1118+3203, and the right panel shows the light curves in the W1 and W2 bands.}
\figsetgrpend

\figsetgrpstart
\figsetgrpnum{2.12}
\figsetgrptitle{RCL AGN_J1412+5400}
\figsetplot{Fig_J1412+5400.pdf}
\figsetgrpnote{The left panel shows the spectra of the RCL AGN J1412+5400, and the right panel shows the light curves in the W1 and W2 bands.}
\figsetgrpend

\figsetgrpstart
\figsetgrpnum{2.13}
\figsetgrptitle{RCL AGN_J1533+0110}
\figsetplot{Fig_J1533+0110.pdf}
\figsetgrpnote{The left panel shows the spectra of the RCL AGN J1533+0110, and the right panel shows the light curves in the W1 and W2 bands.}
\figsetgrpend

\figsetgrpstart
\figsetgrpnum{2.14}
\figsetgrptitle{RCL AGN_J1638+2827}
\figsetplot{Fig_J1638+2827.pdf}
\figsetgrpnote{The left panel shows the spectra of the RCL AGN J1638+2827, and the right panel shows the light curves in the W1 and W2 bands.}
\figsetgrpend

\figsetgrpstart
\figsetgrpnum{2.15}
\figsetgrptitle{RCL AGN_J2146+0032}
\figsetplot{Fig_J2146+0032.pdf}
\figsetgrpnote{The left panel shows the spectra of the RCL AGN J2146+0032, and the right panel shows the light curves in the W1 and W2 bands.}
\figsetgrpend

\figsetgrpstart
\figsetgrpnum{2.16}
\figsetgrptitle{RCL AGN_J2252+0109}
\figsetplot{Fig_J2252+0109.pdf}
\figsetgrpnote{The left panel shows the spectra of the RCL AGN J2252+0109, and the right panel shows the light curves in the W1 and W2 bands.}
\figsetgrpend

\figsetgrpstart
\figsetgrpnum{2.17}
\figsetgrptitle{RCL AGN_J0023+0035}
\figsetplot{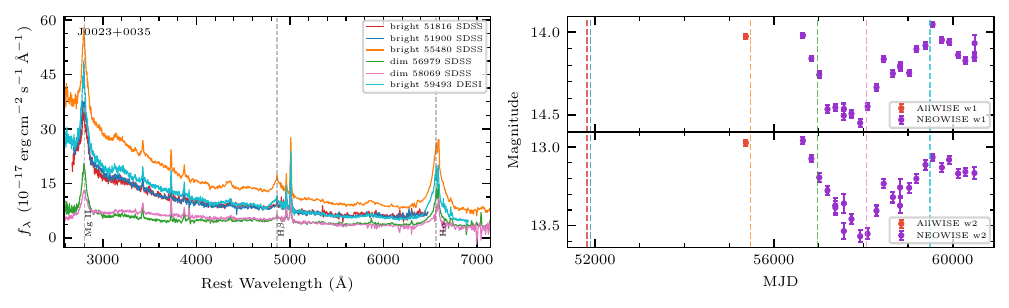}
\figsetgrpnote{The left panel shows the spectra of the RCL AGN J0023+0035, and the right panel shows the light curves in the W1 and W2 bands.}
\figsetgrpend

\figsetgrpstart
\figsetgrpnum{2.18}
\figsetgrptitle{RCL AGN_J1011+5442}
\figsetplot{Fig_J1011+5442.pdf}
\figsetgrpnote{The left panel shows the spectra of the RCL AGN J1011+5442, and the right panel shows the light curves in the W1 and W2 bands.}
\figsetgrpend

\figsetgrpstart
\figsetgrpnum{2.19}
\figsetgrptitle{RCL AGN_J1104+6343}
\figsetplot{Fig_J1104+6343.pdf}
\figsetgrpnote{The left panel shows the spectra of the RCL AGN J1104+6343, and the right panel shows the light curves in the W1 and W2 bands.}
\figsetgrpend

\figsetgrpstart
\figsetgrpnum{2.20}
\figsetgrptitle{RCL AGN_J1324+2802}
\figsetplot{Fig_J1324+2802.pdf}
\figsetgrpnote{The left panel shows the spectra of the RCL AGN J1324+2802, and the right panel shows the light curves in the W1 and W2 bands.}
\figsetgrpend

\figsetgrpstart
\figsetgrpnum{2.21}
\figsetgrptitle{RCL AGN_J1324+4802}
\figsetplot{Fig_J1324+4802.pdf}
\figsetgrpnote{The left panel shows the spectra of the RCL AGN J1324+4802, and the right panel shows the light curves in the W1 and W2 bands.}
\figsetgrpend

\figsetgrpstart
\figsetgrpnum{2.22}
\figsetgrptitle{RCL AGN_J1617+0638}
\figsetplot{Fig_J1617+0638.pdf}
\figsetgrpnote{The left panel shows the spectra of the RCL AGN J1617+0638, and the right panel shows the light curves in the W1 and W2 bands.}
\figsetgrpend

\figsetend

\begin{figure*}
\centering
\includegraphics[width=1\textwidth]{Fig_J0023+0035.pdf}
\caption{The spectra and light curves of the RCL~AGN J0023+0035. The left panel shows the spectra of bright and dim states, where this source exhibits the disappearance followed by the appearance of \hb\ BEL. The right panel shows the light curves in W1 and W2 band, and the six short dotted lines in different colors indicate the epochs of the spectroscopic observations. \\
(The complete figure set, including both spectral and light curves (22 images), is available in the online Journal.)}
\label{Fig_RCLAGN}
\end{figure*}

\section{Results and discussion}\label{sec:discussion}
\subsection{RCL~AGN sample} 
Based on SDSS, LAMOST, and DESI data, supplemented by literature observations, we identify a total of $22$ RCL~AGNs through the combination of quantitative selection and visual inspection. These sources span a wide redshift range ($z=0.022$ to $2.188$) and display RCL behavior in different BELs: $4$ in \ha, $20$ in \hb, $1$ in \Mgii, and $1$ in \Ciiii. Four sources exhibit simultaneous changes in two BELs. In particular, J0043+1344 shows variations in both \Mgii\ and \hb, while J0938+0743, J1011+5442, and J1533+0110 vary in both \hb\ and \ha. The relative BEL variations of these objects are listed in Table~\ref{tab:RCLAGNs}. Among the $20$ RCL~AGNs with available spectra for quantitative BEL measurements, $15$ objects show large BEL variations with both relative variation factors exceeding $0.6$. For three additional objects, one of the two measured variations lies between $0.5$ and $0.6$. Additional two sources exhibit a comparatively modest variation in one transition. The comparatively small $R_{\rm s}$ values in a few cases are likely influenced by the automated spectral fitting. In particular, for low SNR dim spectra, weak residual broad components may be spuriously introduced, even when no genuine BEL is visually apparent. Such residual components lead to a fitted BEL flux in the dim state, thereby reducing the measured relative variation.
 
Among this RCL~AGN sample, J0023+0035, J1104+6343, and J1617+0638 were previously identified as RCL~AGNs by \cite{2025ApJ...981..129W}, J1011+5442, J1324+2802, and J1324+4802 were reported by \cite{2025A&A...693A.173L}, \cite{2025ApJ...986..160D}, and \cite{2024ApJ...966...85Z}, respectively. The remaining $16$ are newly identified in this work, substantially enlarging the known RCL~AGN population.

Within the sample, $19$ objects undergo a turn-off followed by a turn-on transition, whereas three (J0207-0609, J1533+0110, and J2252+0109) exhibit the reverse sequence. The former type is far more numerous than the latter. This predominance of the former type is also noted by \cite{2025ApJ...981..129W}, who identified eight new RCL~AGNs, seven of which undergo a turn-off and later return to a bright state, while only one displays the reverse order. To assess this imbalance, we examined the spectroscopic sampling of our CL~AGN parent sample. Among the $299$ CL~AGNs, $152$ show turn-on transitions and $147$ show turn-off transitions. For the turn-on sources, $33$ have additional spectra obtained before the first CL epoch (${\rm MJD} < {\rm T1}$), $110$ after the second epoch (${\rm MJD} > {\rm T2}$), and $9$ on both sides. The corresponding numbers for the turn-off sources are $26$, $111$, and $10$, respectively. For each object, we measured the longest time interval between the CL-identification spectrum and the most distant additional spectrum, yielding median maximum separations of $\sim$920~days for the turn-on and $\sim$1191~days for the turn-off sources. This indicates that the observed predominance of turn-off followed by turn-on transitions may be influenced by the longer temporal baselines available for monitoring turn-off sources. However, we cannot rule out the possibility that intrinsic differences between the two types of transitions or selection effects may also contribute. Ideally, for a complete sample with a sufficiently long and unbiased observational baseline, the quantities of two types of RC~AGNs are expected to be roughly comparable.

\subsection{Multiple transitions}
We identify $22$ RCL~AGNs among $299$ CL~AGNs, corresponding to an occurrence rate of $\sim 7\%$. This value should be regarded as a lower limit, since some RCL events may have taken place during epochs without spectroscopic coverage or may emerge in the future. This value implies that RCL behavior is not uncommon among CL~AGNs. Previous studies have proposed that CL phenomena may be triggered by disk instabilities \citep{2025ApJS..278...28G,2025A&A...693A.173L,2025ApJ...981..129W}. Once the accretion flow becomes unstable, it may repeatedly switch between different accretion states, leading AGNs to oscillate between bright and dim phases and thus undergo multiple transitions.

\citet{2025A&A...698A.135G} reported that J075947.73+112507.3 exhibits multiple flares in both optical and MIR light curves, corresponding to four to six potential CL events likely driven by accretion rate changes rather than dust obscuration. Similarly, several sources in our sample show multiple flares in the W1 and W2 bands. For example, J1028+2351 exhibits turn-off (from MJD=53734 to MJD=57103) and subsequent turn-on (from MJD=57103 to MJD=58937) transitions, during which the MIR light curves first fade and then brighten. After MJD=58937, both W1 and W2 light curves start to decline again, reaching flux levels comparable to those during the previous dim state (around MJD=57103). This renewed dimming implies that this source may be undergoing again a turn-off transition. In our sample, $14$ sources display similar MIR light-curve patterns, including $13$ objects showing renewed dimming after turn-on transitions and one showing renewed brightening after turn-off transitions. For most CL~AGNs, the maximum variability in the MIR bands exceeds $0.3\,\rm{mag}$ \citep{2022ApJ...927..227L}. Among these $14$ sources, $9$ exhibit strong variability, while the other $5$ display modest changes (listed in Table~\ref{table:multi}). The weaker variability may be due to the fading phase having only recently begun, such that the current observational baseline is insufficient to trace the full changes. These renewed dimming or brightening episodes may represent a third potential CL behavior, which can be further tested with continued photometric and spectroscopic monitoring.

Considering the observed occurrence rate of RCL transitions and the likelihood that confirmed RCL~AGNs may continue to undergo further state changes, we suggest that once an AGN exhibits CL behavior, it may enter a prolonged phase of accretion instability. The currently observed CL and RCL transitions likely represent only short segments within this extended unstable period. Therefore, we speculate that a considerable part of CL~AGNs may experience repeating or multiple transitions, until the accretion flow eventually settles into a stable state. Continued long-term monitoring will help reveal the properties of these accretion instabilities.
 
\begin{deluxetable}{lccccB}
\tablewidth{0pt}
\tablecaption{Multiple CL~AGN candidates \label{table:multi}}
\centering
\tablehead{\colhead{Name} & \colhead{State} & \colhead{MJD} & \colhead{$\bigtriangleup \mathrm{W1}_{max}$} & \colhead{$\bigtriangleup \mathrm{W2}_{max}$}
}
\startdata
J0023+0035	&	dim	&	59553	&	0.12 $\pm$ 0.03	&	0.09 $\pm$ 0.04 \\
J0043+1344	&	dim	&	60131	&	0.27 $\pm$ 0.04	&	0.17 $\pm$ 0.08 \\
J0207-0609	&	bright	&	57404	&	0.47 $\pm$ 0.08	&	0.46 $\pm$ 0.15 \\
J0245+0037	&	dim	&	59590	&	0.14 $\pm$ 0.08	&	0.28 $\pm$ 0.30 \\
J0726+4101	&	dim	&	59619	&	0.34 $\pm$ 0.04	&	0.29 $\pm$ 0.08 \\
J0745+3809	&	dim	&	58770	&	0.25 $\pm$ 0.05	&	0.23 $\pm$ 0.13 \\
J0927+0433	&	dim	&	59900	&	0.20 $\pm$ 0.06	&	0.31 $\pm$ 0.13 \\
J0938+0743	&	dim	&	58960	&	0.16 $\pm$ 0.01	&	0.23 $\pm$ 0.04 \\
J1028+2351	&	dim	&	58966	&	0.42 $\pm$ 0.18	&	0.38 $\pm$ 0.08 \\
J1053+3024	&	dim	&	59541	&	0.49 $\pm$ 0.03	&	0.34 $\pm$ 0.04 \\
J1104+6343	&	dim	&	59314	&	0.51 $\pm$ 0.07	&	0.48 $\pm$ 0.21 \\
J1118+3203	&	dim	&	59697	&	0.36 $\pm$ 0.20	&	0.11 $\pm$ 0.14 \\
J1324+2802	&	dim	&	57751	&	0.72 $\pm$ 0.01	&	0.75 $\pm$ 0.05 \\
J1617+0638	&	dim	&	59634	&	0.31 $\pm$ 0.04	&	0.28 $\pm$ 0.09 \\
\enddata
\tablenotetext{  }{NOTE-Columns: (1) Object name. (2) The latest spectrum state. (3) Epoch for the beginning of brightening or dimming. (4) The maximum magnitude variations of the W1 band after the epoch listed in column (3). (5) The maximum magnitude variations of the W2 band after the epoch listed in column (3). }
\end{deluxetable}

\subsection{Color changes in spectra and MIR light curves}
The multi-epoch spectra in Figure \ref{Fig_RCLAGN} show that, except for prominent BEL variations, the continuum spectral slope changes between bright and dim states. All sources exhibit bluer-when-brighter trends, consistent with previous studies \citep[e.g.,][]{MacLeod2016,Yang2018,Green2022,2024ApJS..270...26G,2025ApJS..278...28G,2025MNRAS.536.2715Z}. This trend is often interpreted as the result of an enhanced accretion rate, which heats the accretion disk and strengthens the UV and optical radiation. In particular, \cite{2025ApJS..278...28G} presented a strong positive correlation between luminosity variations in the continuum and BEL, highlighting the coordinated variations of the disk and broad-line region (BLR). In our sample, the bright spectrum is accompanied by prominent BELs, whereas the dim spectrum shows their disappearance. This coordinated change in the continuum and BELs indicates that the BLR responds to variations in the accretion disk, pointing to accretion rate variations as an important factor in driving CL behaviors.

In addition to variations in spectral color, we examine color changes in the MIR bands. The variability amplitude $m_{\mathrm{var}}$ is estimated following \cite{2002ApJ...568..610E}:
\begin{equation}
m_{\rm var} = \sqrt{S^2-<\sigma^2_{\rm t}>},
\end{equation}
where $S$ denotes the standard deviation of the light curves, $<\sigma^2_{\rm t}>$ is the mean error squared. 
 
The uncertainty $\sigma_{m_{\rm var}}$ is expressed as
\begin{equation}
\sigma_{m_{\rm var}} = \frac{S^2}{m_{\rm var}}\sqrt{\frac{1}{2N}},
\end{equation}
where $N$ is the number of data points. 

\begin{figure*}
\centering
\includegraphics[width=0.5\textwidth]{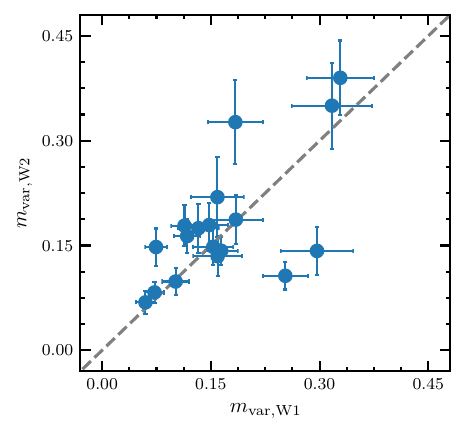}
\caption{Variability amplitudes in the W1 and W2 bands for $18$ RCL~AGNs. The gray dashed line indicates that the variability amplitudes in the W1 and W2 bands are equal. Most sources lie above this line, indicating that the W2 variability amplitude exceeds that of W1.}
\label{Fig:W1_W2}
\end{figure*}

We compute the MIR variability amplitudes during the turn-on phase for sources with more than three valid data points in both the W1 and W2 bands. As shown in Figure \ref{Fig:W1_W2}, most RCL~AGNs exhibit larger variability amplitudes in the W2 band than in W1, indicating a redder-when-brighter trend. This behavior has been widely reported in previous studies \citep[e.g.,][]{Yang2018,2025ApJ...986..160D}. The MIR emission originates from the dusty torus, which absorbs radiation from the accretion disk and re-emits it in the infrared band. This radiation may be influenced by several factors, including the covering factor of hot dust, viewing angle, potential contamination from the host galaxy, and extinction of AGN \citep[e.g.,][]{2008ApJ...685..160N,2016MNRAS.458.2288S}. \cite{2022ApJ...927..107S} reported that this trend is more likely driven by changes in the covering factor of hot dust, rather than host galaxy contamination or viewing angle. \citet{2017MNRAS.472.3492E} suggested that the covering factor tends to decrease with increasing Eddington ratio. Therefore, the observed redder-when-brighter trend likely traces structural and radiative variations within the dusty torus.
 
Two sources, J1104+6343 and J1118+3203, display the opposite behavior, with larger amplitudes in W1 than in W2. The physical origin of this deviation is not yet clear, though we cannot rule out the possibility that it is caused by measurement uncertainties. Alternatively, it might hint at differences in dust composition or geometry in these systems, which would require further observations to confirm.

\subsection{Black hole mass and Eddington ratio estimates}

The black hole mass ($M_{\rm BH}$) is one of the most fundamental parameters of AGNs, and the Eddington ratio is currently believed to play a key role in determining the occurrence of CL phenomena. In this subsection, we describe how we estimate $M_{\rm BH}$ and the Eddington ratios for our sample.

\subsubsection{Methodology for black hole mass estimation}
All black hole mass measurements in this work are estimated using single-epoch (SE) spectra based on the empirical radius–luminosity ($R$–$L$) relation for the BLR. The virial mass is computed as
\begin{equation}
    M_\mathrm{BH} = f \frac{R_\mathrm{BLR} {\Delta V}^2}{G},
\end{equation}
where $\Delta V$ is the width of a BEL (typically characterized by the full width at half maximum, FWHM), $R_{\rm BLR}$ is inferred from the $R$–$L$ relation, $G$ is the gravitational constant, and $f$ is the virial factor (we adopt $f=2.6$ following convention for SE masses; see \citealt{2015ApJ...809..123H}).

The BLR radius is derived from the continuum luminosity at a given wavelength according to empirical relations calibrated by reverberation mapping:
\begin{equation}
R_{\rm BLR} = \alpha \left( \frac{L_{\lambda}}{10^{44}\, {\rm erg\,s^{-1}}} \right)^{\beta} \, {\rm lt\text{-}days},
\end{equation} 
where $L_{\lambda}$ is the continuum luminosity at rest frame wavelength, with $L_{5100}(z \leq 0.7)$ and $L_{3000}(z > 0.7)$. The scaling parameters ($\alpha$, $\beta$) depend on the choice of rest frame wavelength $\lambda$, with $\alpha=24.5\pm1.8$ and $\beta=0.608\pm0.045$ for $L_{5100}$, $\alpha=25.2\pm3.0$ and $\beta=0.47\pm0.05$ for $L_{3000}$ \citep{2005ApJ...629...61K,2004MNRAS.352.1390M}.

With $M_{\rm BH}$ determined, the corresponding Eddington luminosity $L_{\rm Edd} = 1.26 \times 10^{38}({M_{\rm BH}}/{M_{\odot}}) \, {\rm erg\,s^{-1}}$ can be derived (see \citealt{1984ARA&A..22..471R}). The Eddington ratio is then simply $\lambda_{\mathrm{Edd}} = L_{\mathrm{bol}}/L_{\mathrm{Edd}}$, where the bolometric luminosity $L_{\rm bol}$ is driven from the continuum luminosity using standard bolometric corrections (e.g., $L_{\rm bol} = 9\, L_{5100}$ from \citealt{2000ApJ...533..631K}; $L_{\rm bol} = 5.15\,L_{3000}$ from \citealt{2006ApJS..166..470R}). 

\subsubsection{Adopted data sources}

To ensure reliable measurements of the black hole mass, we preferentially used spectra obtained during the high states of the RCL~AGNs, when both the continuum and BELs are stronger. For RCL~AGN exhibiting a turn-off followed by a turn-on transition, the spectrum at $\rm{MJD_1}$ is used to estimate black hole mass; for those showing a turn-on followed by a turn-off, the spectrum at $\rm{MJD_2}$ is used to estimate black hole mass. For $21$ sources ($z\leq0.7$), we measured FWHM of \hb\ and $L_{5100}$ from spectral fitting to estimate back hole mass. For the one high redshift source with variable \Ciiii\ emission line, robust SE mass calibrations are not yet firmly established. We defer detailed mass estimates for these objects to future work, but note that possible approaches include using (i) \Ciiii\ scaling relations with corrections for non-virial contributions (e.g., \citealt{2017MNRAS.465.2120C}), or (ii) empirical correlations between \Ciiii\ and H$\beta$/\Mgii\ luminosities. 

\subsubsection{Distribution of Eddington ratios}
We estimated the black hole masses of our sample sources as far as possible using SE spectra. Specifically, $21$ sources were estimated based on the broad \hb\ line. In addition, the black hole mass for one source was not estimated, as it exhibits RCL behavior in \Ciiii, for which no reliable virial calibrations for this line.

Most of these objects have at least three spectroscopic epochs, having undergone two state transitions. For each object, we selected spectra representative of the different states to calculate the corresponding $\lambda_\mathrm{Edd}$. The derived $\lambda_\mathrm{Edd}$ and their observation dates are listed in Table \ref{tab:RCLAGNs}. Among the $21$ sources with black hole mass estimates, three experienced a turn-on followed by a turn-off sequence, and two have dim spectra taken from the published figure rather than from a public spectroscopic survey. These five objects were excluded, leaving $16$ sources in the final subsample used for Eddington ratio analysis.

\begin{figure*}
\centering
\includegraphics[width=0.5\textwidth]{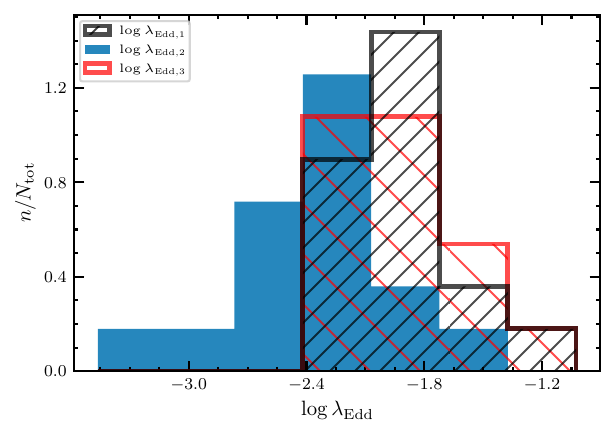}
\caption{Distributions of Eddington ratio for $16$ RCL~AGNs that exhibit the turn-off transition followed by the turn-on transition. The black, blue, and red histograms represent the Eddington ratio derived from spectra obtained at MJD1, MJD2, and MJD3, respectively. The spectra at MJD1 and MJD3 correspond to the bright states, while the spectrum at MJD2 corresponds to the dim state. These distributions exhibit the changes in Eddington ratios across the bright and dim phases.}
\label{Fig_L_R}
\end{figure*}

Figure \ref{Fig_L_R} shows the distribution of Eddington ratios for these $16$ sources. All of them experienced a ``turn-off–turn-on'' transition sequence. In the figure, different colors denote the Eddington ratios at different states. The Eddington ratios in the low state range from about $0.004$ to $0.094$, while those in the two high states are comparable, typically between $0.0003$ and $0.03$. Statistically, the Eddington ratios differ by more than a factor of $3$ between the high and low states. Such a large variation over a relatively short timescale strongly suggests that the accretion disk underwent a state transition.

Previous studies have emphasized the importance of the accretion rate in driving CL behavior. \citet{2019ApJ...874....8M} found that CL~AGNs have lower Eddington ratios than extremely variable quasars and less variable quasars, suggesting that CL behavior is more likely to occur at lower Eddington ratios. More recent works using different CL~AGN samples \citep[e.g.,][]{Green2022,2024ApJ...966..128W,2025ApJS..278...28G,2025ApJ...986..160D} reached similar conclusions. In particular, \citet{2024ApJ...966..128W} reported that transitions occur around $\log(L_{\rm bol}/L_{\rm Edd}) \sim -2$. \citet{2019ApJ...883...76R} proposed an analogy between CL~AGNs and X-ray binaries, both of which exhibit a V-shaped inversion between the UV-to-X-ray spectral index ($\alpha_{\rm OX}$) and the Eddington ratio. This inversion occurs in a critical Eddington ratio of $\sim 0.01$, marking a transition between different accretion states. The Eddington ratio distributions of our sample are consistent with these studies. In particular, the median Eddington ratio in the dim state lies close to this critical value, supporting the view that RCL transitions are driven by changes in the accretion rate and preferentially occur at the critical Eddington ratio.

\subsection{MIR variability timescale in CL behavior}
The timescale is a key quantity in constraining the physical models of CL~AGNs. For instance, whether the turn-off and turn-on timescales are symmetric, and whether the timescale correlates with other physical parameters (such as black hole mass, Eddington ratio, or luminosity). 

To characterize the timescale associated with the largest photometric variation during the CL phase, we primarily rely on photometric light curves rather than the sparsely sampled spectroscopic observations. Owing to their much denser temporal coverage, photometric data can provide significantly tighter constraints on the transition timescale. Among the available photometric datasets, we find that the optical light curves often exhibit relatively large scatter, and in some cases the variability features are not clearly discernible. This is likely caused by a combination of host-galaxy contamination, seeing variations, and the fact that some sources approach the limiting magnitude of the ZTF survey, which increases the photometric uncertainties. In contrast, the MIR light curves reveal cleaner long-term variability structures with substantially reduced scatter. Moreover, the WISE observations generally provide better temporal coverage for most sources in our sample, making them more likely to capture both the bright and faint states of the variability evolution. Therefore, we adopt the WISE MIR light curves to characterize the variability timescales associated with the CL phase.

RCL events offer unique opportunities to more robustly constrain the timescale of the luminosity evolution. The light curves of RCL~AGNs often display more complex structures, with multiple local maxima and minima associated with repeated transitions. These features allow us to identify the boundaries of high and low states and thus extract the corresponding variability timescales. Our approach is as follows. First, we determine the spectral high and low states of each source based on spectroscopic observations. Owing to the sparse cadence of the spectroscopic observations, the epochs of the high- and low-state spectra generally do not coincide with the maxima and minima of the WISE light curves. To account for this, we identify the nearest local maximum and local minimum in the MIR light curves around the epochs of the high- and low-state spectra, respectively. Next, we check whether additional significant transitions occur between these two local extrema. If no further transitions are evident, we define the MIR variability timescale as the temporal separation between the nearest local maximum and the local minimum. If multiple transition-like features are present in the interval, we instead identify the closest pair of adjacent transitions and adopt the time difference between them as the MIR variability timescale. For most objects, the WISE light curves during the intervals between two spectroscopic epochs do not exhibit complicated structures, making the determination of MIR variability timescales relatively straightforward.

In a few cases, the MIR light curves do not cover the full transition or lack local extrema near the spectroscopic epochs. For such sources, we instead estimate a lower limit. We first search for the nearest local extrema in the MIR light curves around the epochs of the spectra. If no local extrema exist, we use the photometric data point closest in time to the corresponding spectrum to represent that state. The MIR variability timescales is then defined as the time interval between the two selected MIR epochs. Because the full transition may not be captured, the resulting timescale should be regarded as a lower limit.  
%%%%%%%%%%%%%%%%%%%%%%%%%%%%%%%%%%%%%%%%%%%%%%%%%%

\begin{figure*}
\centering
\includegraphics[width=0.5\textwidth]{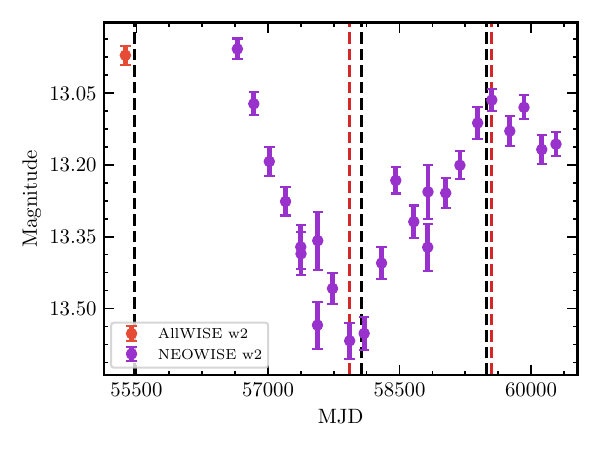}
\caption{
The MIR light curves of J0023+0035. The black dotted lines mark the epoch of spectral observations, corresponding to $\mathrm{MJD_1}$, $\mathrm{MJD_2}$, and $\mathrm{MJD_3}$. The red dotted lines indicate the WISE-determined epochs.
} 
\label{Fig_MJD}
\end{figure*}

The WISE mission has been monitoring since 2010, whereas the earliest spectra of many sources were obtained much earlier (see Figure \ref{Fig_MJD}). As a result, the WISE light curves generally do not cover the first transition event. In addition, J1638+2827 shows no significant MIR variability because their MIR brightnesses approach the WISE sensitivity limit, and hence no measurable MIR variability timescale can be derived for them. After excluding this object, we derive the MIR variability timescales during the second transition for $21$ sources, among which J1118+3203 has lower limits of the MIR variability timescale due to the lack of simultaneous local maxima and minima near the spectroscopic epochs. The MIR variability timescales derived for the $21$ RCL~AGNs are summarized in Table~\ref{tab:RCLAGNs}. Among them, three objects display dimming behavior within the WISE coverage. For the sake of consistency and to avoid mixing potentially different timescale definitions, we restrict our analysis to the brightening behavior when deriving the statistical properties of MIR variability timescale. Based on our statistical analysis, the MIR variability timescales during the brightening phase in the rest frame range from about $2$ to $5.3$ years.

The MIR emission originates from dust reprocessing of the central optical/UV radiation and may be affected by the covering factor and dust torus geometry. These factors affect the observed MIR variability in two aspects: (1) short-timescale variability features are smoothed out on variability timescales shorter than the characteristic spread of the dust response, and (2) the MIR-optical time delays in the bright and faint states are different, introducing additional dispersion into the MIR-derived variability timescales. The dispersion can be quantified by the difference in the optical-to-MIR time delay between the bright and faint states, $\Delta\tau = |\tau_{\rm bright} - \tau_{\rm faint}|$, as the geometric spread of the dust torus is smaller than the change in the mean reverberation lag between the two accretion states. Here, $\tau_{\rm bright}$ and $\tau_{\rm faint}$ are estimated using the empirical $R_{\rm dust}$--$L_{\rm bol}$ relation from \citet{2024ApJ...968...59M}. The median value of $\Delta\tau$ is approximately $51$ days, and the MIR variability timescale is more than eight times larger than $\Delta\tau$ for most sources. This indicates that dust reprocessing introduces modest dispersion relative to the several-year MIR variability timescales derived in this work. Therefore, although the MIR-derived timescales should not be interpreted as direct measurements of the intrinsic accretion-disk transition timescale, they can still serve as useful empirical tracers of the long-term CL variability evolution driven by changes in the central radiation field. Motivated by this, we further explore whether these MIR variability timescales correlate with black hole mass.

%%%%%%%%%%%%%%%%%%%%%%%%%%%
\begin{figure*}
\centering
\includegraphics[width=0.5\textwidth]{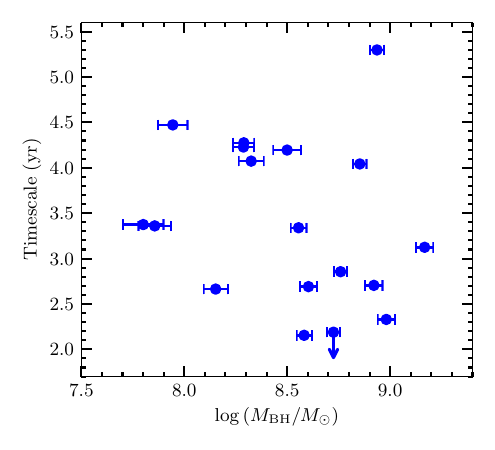}
\caption{Distribution of black hole mass and MIR variability timescales during the brightening phase} for $18$ RCL~AGNs.
\label{Fig_Mbh_MJD}
\end{figure*}

We aim to test the possible connection between the MIR variability timescales during the brightening phase and black hole masses (Figure~\ref{Fig_Mbh_MJD}). The correlation analysis between the two quantities yields Pearson and Spearman's rank coefficients of $-0.212$ and $-0.311$ with corresponding $p$-values of $0.398$ and $0.210$, respectively, suggesting no significant correlation between the two quantities. 

Theoretically, the viscous timescale of the SSD model increases with the black hole mass. Observational studies of AGN variability also reported a positive correlation between the variability timescale (which is suspected to be related to the thermal timescale) and the black hole mass \citep[e.g.,][]{2019ApJ...877...23L,2021Sci...373..789B,2024ApJ...966....8Z}. If the CL behaviors are driven by thermal or viscous instabilities, the MIR variability timescale is expected to increase with the black hole mass. This expectation is disfavored by our analysis (Figure~\ref{Fig_Mbh_MJD}). Several factors may contribute to this inconsistency. First, our sample size remains relatively small, which may not be sufficient to robustly detect the underlying correlation. Second, selection effects may play a role, as the MIR variability timescales during the brightening phase in our sample are clustered within a narrow range of approximately $2$-$5$ years, likely representing only a small fraction of the full distribution and thus obscuring the intrinsic correlation. Third, the MIR variability timescales carry additional dispersion from dust reprocessing, while the black hole masses estimated from single-epoch virial methods also carry substantial uncertainties. The scatter in both quantities may together obscure the underlying correlation. Therefore, the absence of a detected correlation between the MIR variability timescale and black hole mass does not imply the absence of an intrinsic relation between the CL timescale and black hole mass. Future studies based on larger CL AGN samples and densely sampled optical monitoring will provide more direct constraints on the accretion-disk variability timescales, enabling a more stringent test of a possible connection between black hole mass and the CL transition timescale. 

\section{Summary}\label{sec:summary}

In this work, we compiled a large sample of $1,154$ CL~AGNs from the literature and collected their available spectra from SDSS, LAMOST, and DESI. Among them, $299$ sources have additional spectra beyond those used for the initial CL confirmation, allowing us to search for repeated type transitions. The identification of RCL~AGNs was performed through spectral fitting-based selection and visual inspection. We further collected MIR light curves from WISE for all objects to verify whether repeated changing-look behaviors occurred. We then conducted analyzes on the selected samples and obtained the following results.

\begin{itemize}
\item We identified $22$ RCL~AGNs, which exhibit RCL behaviors in \ha, \hb, \Mgii, or \Ciiii\ emission lines. Of these, $19$ exhibit a turn-off followed by a turn-on transition, while $3$ show the reverse sequence. In total, $16$ are newly identified. Compared with the $28$ RCL~AGNs previously reported in the literature \citep{2024ApJ...970...85W,2025ApJ...981..129W,2025A&A...693A.173L}, our work significantly expands the known RCL~AGN sample.
\item The $22$ RCL~AGNs are identified from the $299$ CL~AGNs, corresponding to a lower-limit occurrence rate of $\sim7\%$. Among these, $14$ sources show renewed MIR variability following a second bright or dim state, indicating a potential third CL event. Such multiple CL behaviors are likely driven by accretion state instabilities and we suspect that a considerable part of CL~AGNs may undergo repeating or multiple transitions.
\item We estimated black hole masses and derived Eddington ratios in dim and bright states. The dim state exhibits a lower Eddington ratio compared to the bright state, characterized by a median value of $\sim0.01$. This result is consistent with previous studies and supports the idea that RCL behavior is driven by the change in the Eddington ratio.
\item The MIR variability timescales during the brightening phase of $18$ RCL~AGNs are well constrained by spectra and densely sampled MIR light curves. Our result shows no significant correlation between the black hole mass and the MIR variability timescale during the brightening phase, which differs from the positive correlation expected from accretion theories. This result may be tentative due to the limited sample and observational limitations.
\end{itemize}

\section{Acknowledgments}
We thank the anonymous referee for constructive comments that improved the paper. This work was supported by the National Key R\&D Program of China under grants 2023YFA1607900 and 2023YFA1607903, the National Natural Science Foundation of China under grants 12533006, 12433007, 12221003, 12263003, 12322303, 12503019, 12263003 and 12363002, and the ResearchStart-up Foundation of Quanzhou Normal University (H25052).

This work makes use of public data from the SDSS, LAMOST, DESI, WISE, and ZTF. LAMOST, also known as the Guoshoujing Telescope, is a National Major Scientific Project built by the Chinese Academy of Sciences, with funding provided by the National Development and Reform Commission, and is operated and managed by the National Astronomical Observatories, Chinese Academy of Sciences. Funding for SDSS has been provided by the Alfred P. Sloan Foundation, the Heising-Simons Foundation, the National Science Foundation, and participating institutions. The DESI Legacy Imaging Surveys consist of the DECaLS, BASS, and MzLS surveys conducted using the Blanco, Bok, and Mayall telescopes. WISE is a joint project of the University of California, Los Angeles, and the Jet Propulsion Laboratory/California Institute of Technology, funded by the National Aeronautics and Space Administration, and NEOWISE is a project of the Jet Propulsion Laboratory/California Institute of Technology funded by the NASA Planetary Science Division. ZTF is supported by the National Science Foundation and a collaboration of participating institutions. We thank the respective teams and funding agencies for making these data publicly available. For detailed acknowledgments and funding information, please refer to the original publications and data release notes.

%%%%%%%%%%%%%%%%%%%%%%%%%%%%%%%%%%%%%%%%%%%%

\bibliography{ref}{}
\bibliographystyle{aasjournal}

\end{document}